\documentclass[a4paper,11pt]{article}
\pdfoutput=1 % if your are submitting a pdflatex (i.e. if you have
\usepackage{jinstpub} % for details on the use of the package, please
\usepackage{titlesec}                     % see the JINST-author-manual
\usepackage{parskip}
\usepackage{float}
\usepackage{bold-extra}
\usepackage{hyperref}

\usepackage{times, txfonts}
\titlespacing{\section}{0pt}{\parskip}{2pt}
\titlespacing{\subsection}{0pt}{\parskip}{0pt}
\titlespacing{\subsubsection}{0pt}{\parskip}{-\parskip}

\author[m,1]{M.I. Pedraza-Morales,\note{Corresponding author.}}
\author[i,1]{M.A. Shah,}
\author[u,1]{M.Shopova}

\affiliation[i]{National Centre for Physics, Quaid-i-Azam University, Islamabad, Pakistan}
\affiliation[m]{Benem\'erita Universidad Autonoma de Puebla, Puebla, Mexico}
\affiliation[u]{Bulgarian Academy of Sciences, Inst. for Nucl. Res. and Nucl. Energy, Tzarigradsko shaussee Boulevard 72, BG-1784 Sofia, Bulgaria}

\title{\boldmath First results of CMS RPC performance at 13 TeV}

\emailAdd{mpedraza@fcfm.buap.mx, mashah@cern.ch, maryana.vutova@cern.ch}

\abstract{The muon spectrometer of the CMS (Compact Muon Solenoid) experiment at the Large Hadron Collider (LHC) is equipped with a redundant system made of Resistive Plate Chambers (RPCs) and Drift Tube (DT) chambers in the barrel, RPC and Cathode Strip Chambers (CSCs) in the endcap region. In this paper, the first results of the performance of the RPC system during 2015 with the LHC running at 13 TeV is presented. The stability of the RPC performance, in terms of efficiency, cluster size and noise, is reported. %Finally, the radiation background levels on the RPC system have been measured as a function of the LHC luminosity. Extrapolations to the High Luminosity LHC conditions are also discussed.
}


\arxivnumber{} % only if you have one

\collaboration[c]{ on behalf of the CMS Collaboration}
\proceeding{$13^{th}$ Workshop on Resistive Plate Chambers and Related Detectors\\
  February 22-26, 2016\\
  Ghent University, Belgium}

\newcommand{\fig}{figures/}
\usepackage{graphicx, subfigure}

\begin{document}
\maketitle
\flushbottom

\section{Introduction}
\label{sec:intro}
Muons provide a clean signal to detect interesting events over complicated backgrounds at the LHC~\cite{Evans:2008zzb}. The muon system in the Compact Muon Solenoid (CMS) experiment ~\cite{Chatrchyan:2008aa} has the following primary functions: muon triggering, transverse momentum measurement, muon identification and charge determination. The muon system allows to identify the muons produced in many Standard Model processes, like top quark, W and Z decay, and the well-known Higgs Boson on top of them \cite{Aad:2015zhl}. Hence a robust and redundant muon spectrometer is needed to provide efficient muon reconstruction.

\begin{figure}[!htb]
\centering
\includegraphics[width=.8\textwidth]{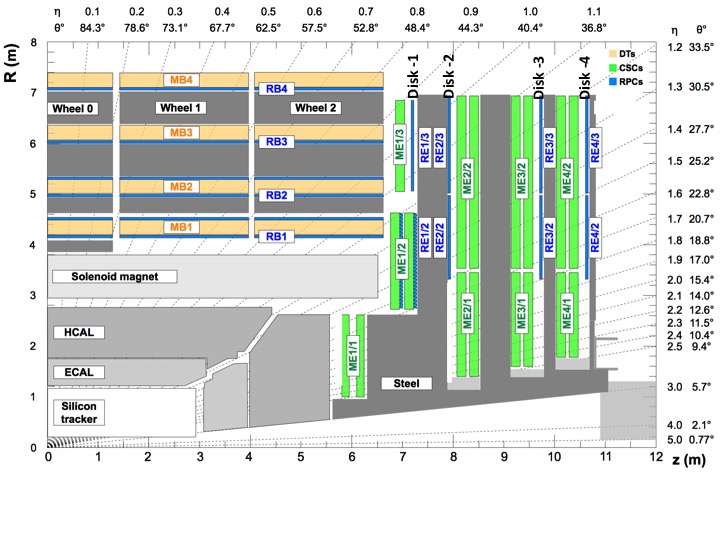}
\label{fig:figure0}
\caption{A quadrant of the muon system, showing DT chambers (yellow), RPC (light blue), and CSC (green). }
\end{figure}

The muon system, described in detail in ~\cite{CMS:1997dma}, uses three different technologies; drift tubes (DT) in the barrel region, cathode strip chambers (CSC) in the endcap region, and resistive plate chambers (RPC) in both the barrel and endcap, and it covers a pseudorapidity region $ |\eta| < 2.4$. The DTs and RPCs in the barrel cover the eta region $ |\eta| < 1.2$, while the CSCs and the RPCs in the end-caps cover the eta region $0.9 < |\eta| < 2.4$.  During the first long shutdown (LS1) of the LHC (2013-2014) tfhe CMS muon system was upgraded with 144 RPCs on the endcap (Disk$\pm$4). This addition increased the overall robustness of the CMS muon spectrometer and improved the muon reconstruction efficiency, in the range $1.2 < |\eta| < 1.8$. Figure \ref{fig:figure0} shows a longitudinal layout of one quadrant of the muon spectrometer, where (0,0) is the collision point.

For the RPC system the barrel region is divided into 5 separate wheels (named Wheel$\pm 2$, Wheel$\pm 1$ and Wheel-0) while the endcaps in 4 disks both in the forward and backward directions (named Disk$\pm 4$, Disk$\pm 3$, Disk$\pm 2$, Disk$\pm 1$). Each wheel is divided into 12 sectors while every disk into 36 sectors, corresponding to $\phi$ partitions in the plane perpendicular to the beam pipe. In total there are 1056 RPC chambers, covering an area of about 3950 $m^2$, equipped with about 123,688 readout strips. The CMS RPCs are double-gap chambers with 2 mm gas width each and copper readout in between. The bakelite bulk resistivity is in the range of $2-5$ $10^{10} \Omega cm$. They operate in avalanche mode with a gas mixture of 95.2\% $C_2H_2F_4$, 4.5\% $iC_4H_{10}$ and 0.3\% $SF_6$.

\section{RPC Performance during RUN-1 and RUN-2}

\subsection{RPC Working Point Calibration}

Dedicated runs are taken to perform high voltage (HV) scans at least once per year at the beggining of data taking. Collision data is recorded at several HV settings during a series of runs to define the optimal operating voltage for each chamber, called working point (WP). Details can be found in ~\cite{e, f} for a full explanation of the HV scan, dependence of efficiency on the HV, including the analysis and methodology. The dependence of the avalanche production on the environmental pressure P, temperature T and the applied HV can be summarised in an effective HV equation \eqref{eq:y:3}.
\begin{equation}
\label{eq:y:3}
HV_{eff}(p,T) = HV_{app} (p_0/p)(T/T_0)
\end{equation}
Where HV\textsubscript{eff} \cite{Abbrescia:2013eua} is effective high voltage, HV\textsubscript{app} is applied high voltage, and the reference temperature and pressure are T\textsubscript{0} = 293 K and p\textsubscript{0} = 965 mbar (average pressure in the cavern).  

\begin{figure}[!htb]
\centering
%\subfigure[]{%
\includegraphics[width=.7\textwidth]{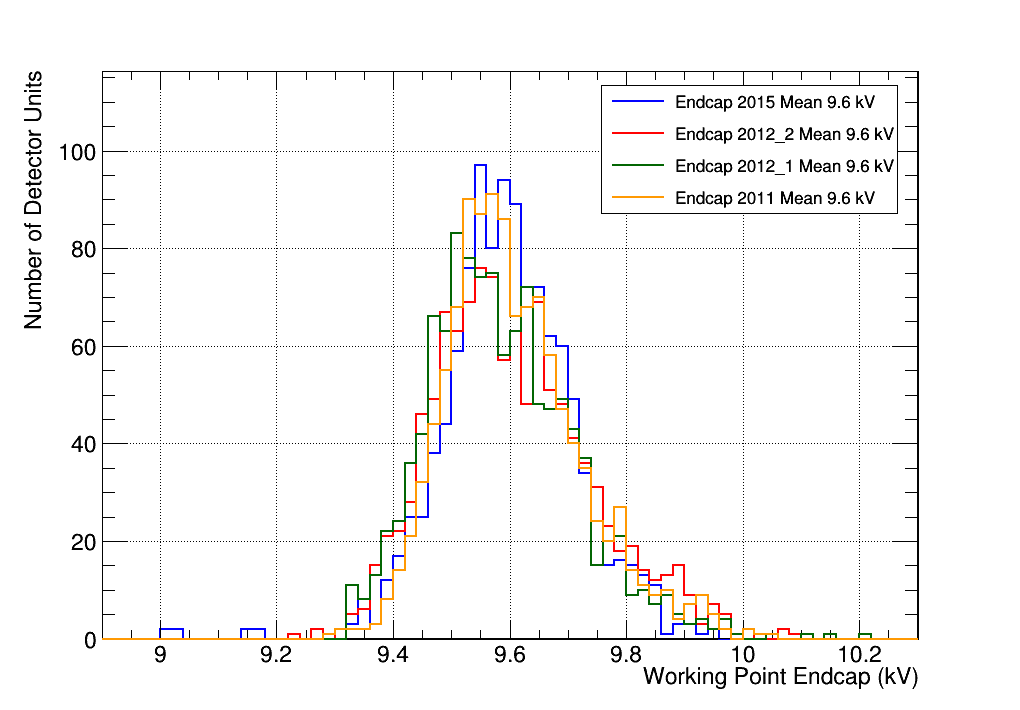}
%\label{fig:subfigure1}}
%\quad
%\subfigure[]{%
%\includegraphics[width=.8\textwidth]{\fig/z2.png}
%
\caption{ HV\textsubscript{eff} distributions of the WP's for the encap as measured at the 95$\%$ of the efficiency for the same runs as for the working point.}
%\caption{Barrel and endcap mean working points through the LHC working years, the error bars correspond to the sigmas obtained from }Ωthe fit to the distribution. From the plot there is no significant differences in the mean values and thus no significant ageing effect is observed. }
\label{fig:figure9a}
\end{figure}

The WP is defined as the HV at which a chamber reaches 95$\%$ of the plateau efficiency plus 100 V for the barrel and 120 V for the endcap. This difference allows barrel and endcap to provide similar global efficiency. The difference correspond to the fact that the endcap has more layers than the barrel, thus working at higher HV the barrel is able to provide a compatible efficiency. Figure \ref{fig:figure9a} shows no significative variations on the working point over the years.

\begin{figure}[!htb]
\centering
%\subfigure[]{%
%\includegraphics[width=.7\textwidth]{\fig/z1.png}
%\label{fig:subfigure1}}
%\quad
%\subfigure[]{%
\includegraphics[width=.7\textwidth]{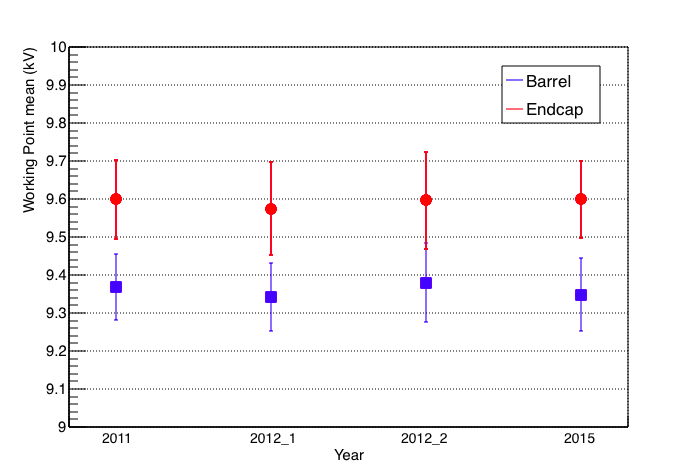}
%
%\caption{In (a) the HV\textsubscript{eff} distributions of the WP's for the encaps as measured at the 95$\%$ of the efficiency for the same runs as for the working point. 
\caption{The core of the WP distributions, shown in Figure\ref{fig:figure9a}, have been fitted to Gaussian. The points on plot correspond to the mean values and the error bars to the sigmas obtained from the fit. The values in the Barrel are shown in blue and the Endcap in red.  }
\label{fig:figure9}
\end{figure}

%The X-axis corresponds to the phi partitions of each wheel, 12 sectors per wheel, while the Y-axis corresponds to the partitions of the detector.
The distributions for 2011, 2012 and 2015 as shown in figure \ref{fig:figure9} show no evident ageing effect. The high voltage shift between barrel and endcap chambers depends on few differences in the assembly parameters and the definition of the working point.

\subsection{RPC Background}
Background radiation level in the CMS muon system is one of the important factors in the overall performance. Low-momentum primary and secondary muons, low-energy gamma-rays, neutrons, and LHC beam-induced backgrounds could have an impact on performance of trigger and pattern recognition of muon tracks. In addition, excessive radiation levels can also cause premature ageing of the detectors.
\begin{figure}[!htb]
\centering
%\subfigure[]{%
\includegraphics[width=.49\textwidth, height=7cm ]{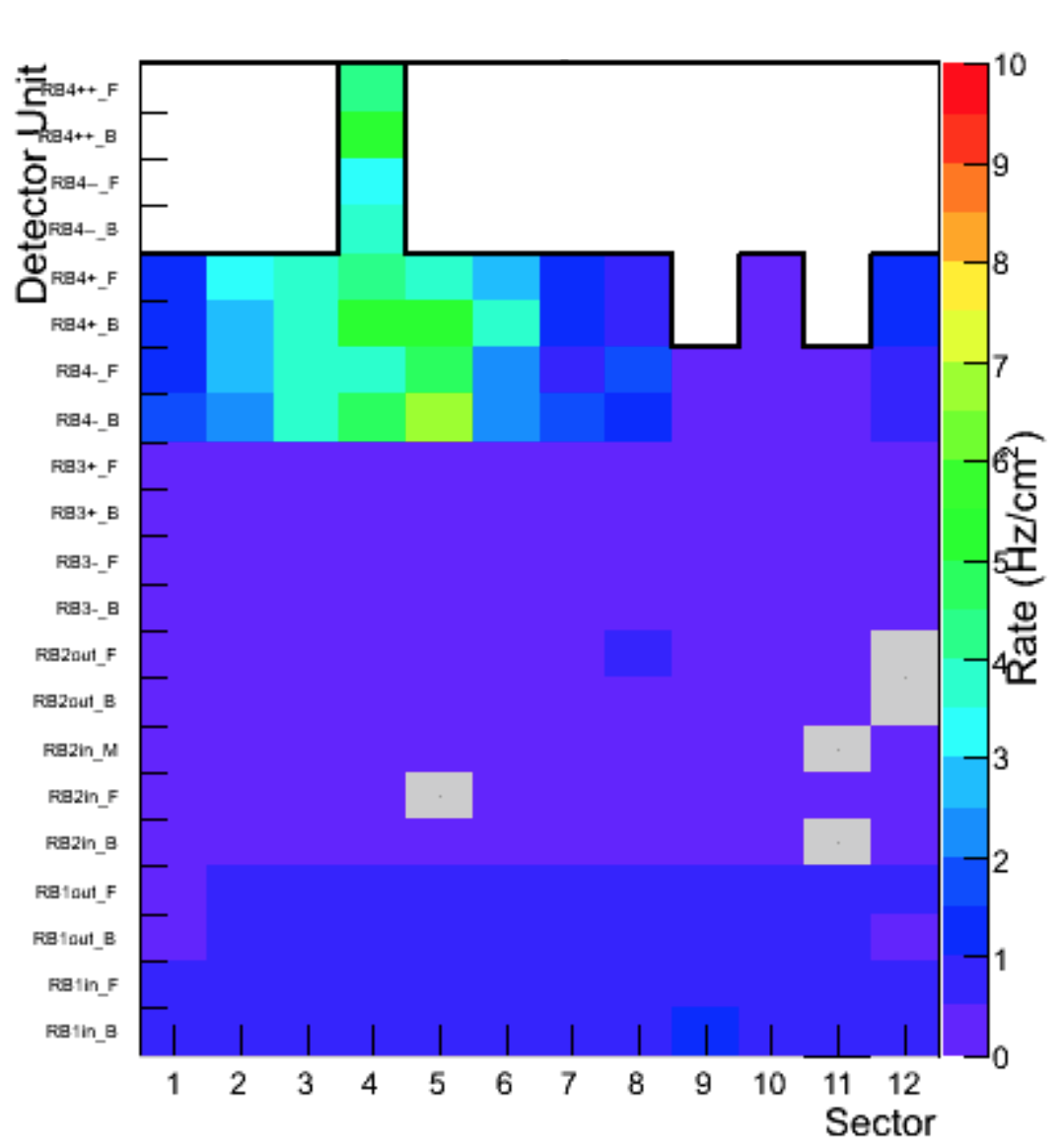}
%\label{fig:subfigure1}%}
%\quad
%\subfigure[]{%
\includegraphics[width=.49\textwidth, height=7cm ]{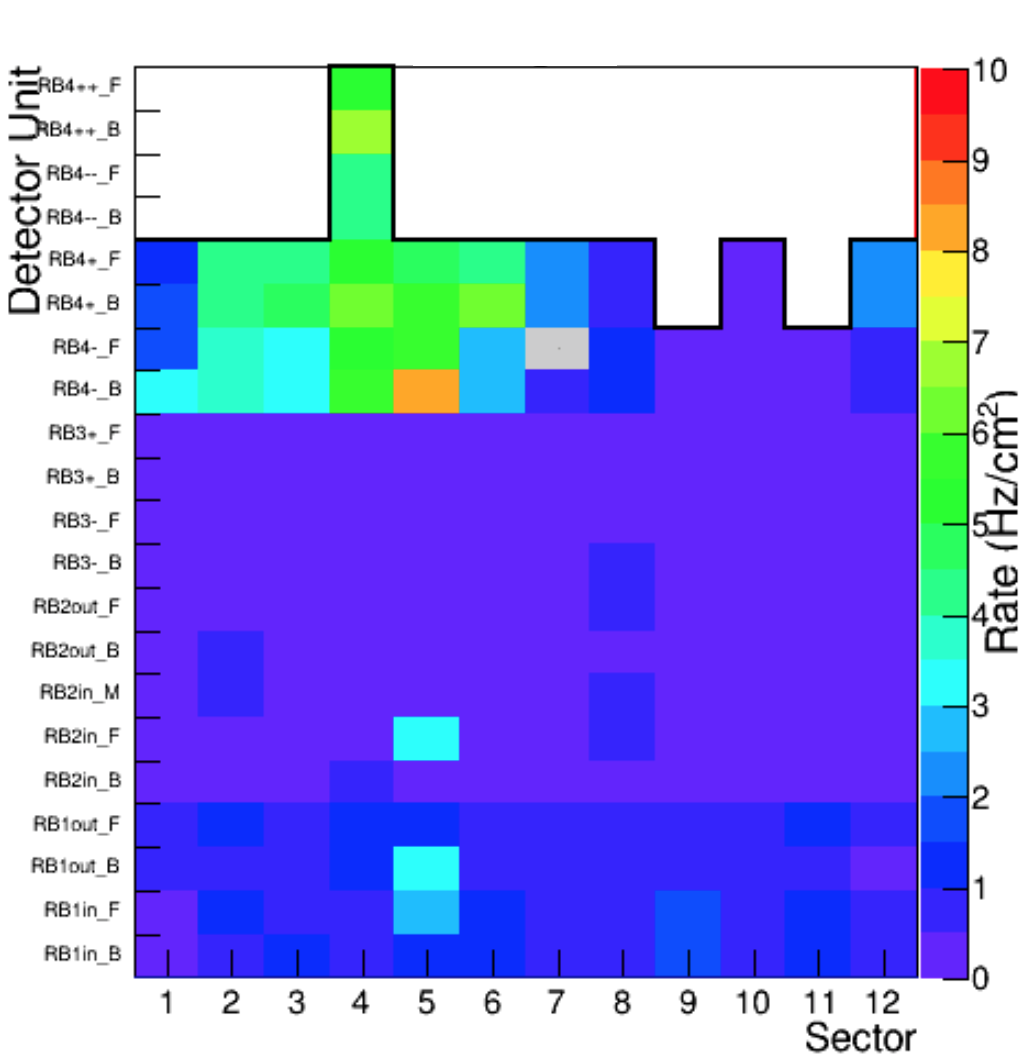}
%\label{fig:subfigure2} %}
%
\caption{The detector units hit rate (in Hz/cm\textsuperscript{2}) is shown for a run at average instantaneous  luminosity of 4.5*10\textsuperscript{33} cm\textsuperscript{-2}s\textsuperscript{-1}  for Wheel-0 at 8 TeV before 2013 (left) and at 13 TeV during 2015 (right). Detector units switched off are shown in gray. Blue and violet colours  correspond to lower rates, while yellow, orange and red colours correspond to high background level. The X-axis corresponds to the sector, 12 per wheel, while the Y-axis corresponds to the eta partitions of the chamber. }
\label{fig:figure12345}
\end{figure}

The main contribution to the RPC rate is coming from radiation background. Figure~\ref{fig:figure12345} shows an example of 2015 at 13 TeV and its corresponding results with 2012 at 8 TeV. The rate is homogeneously distributed except in top sectors (3,4,5) of outermost stations (RB4). This effect is due to the thermo-neutrons populating the cavern during the LHC operation.  

\begin{figure}[!htb]
\centering
%\subfigure[]{%
\includegraphics[width=.49\textwidth]{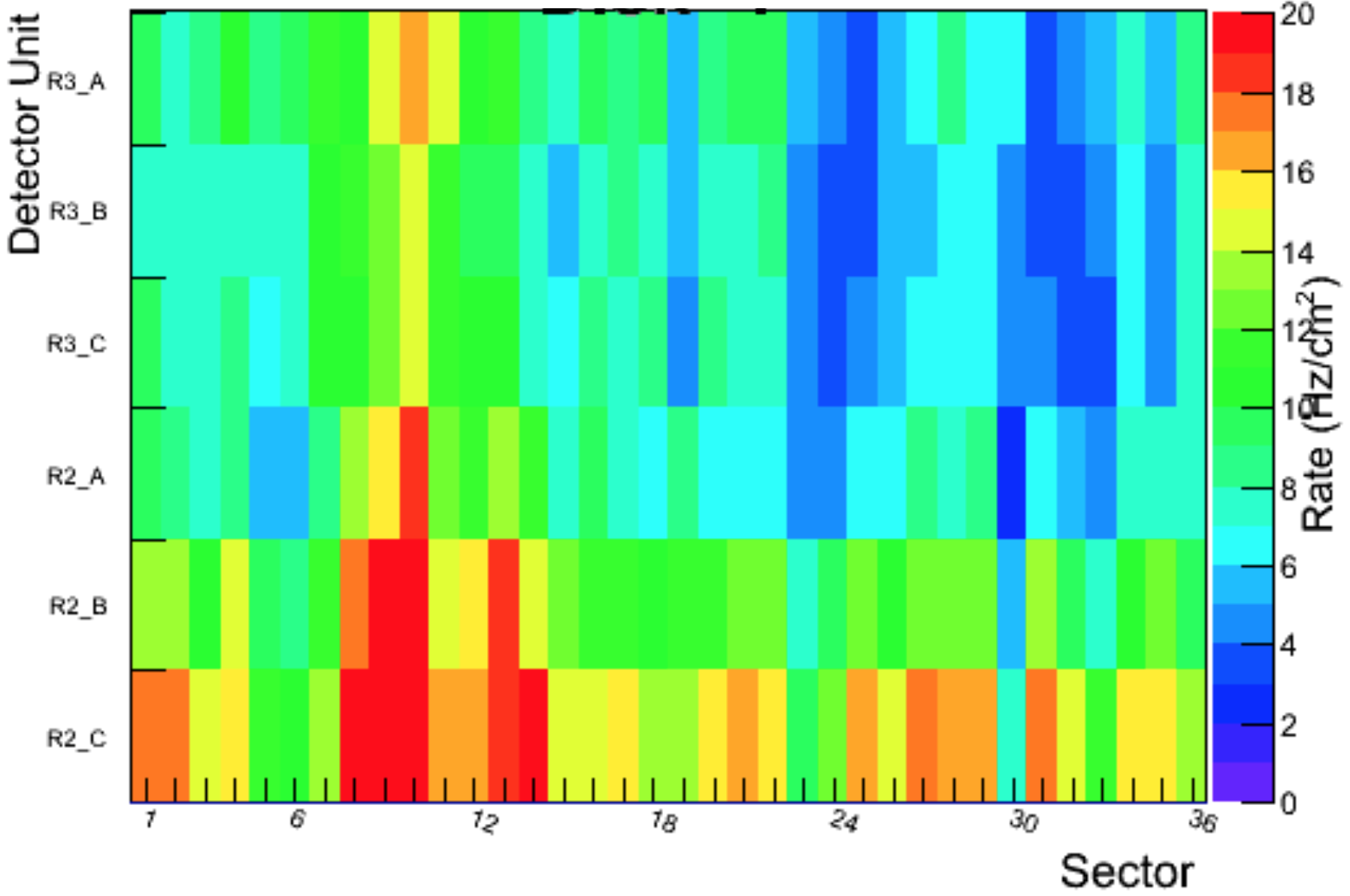}
%\label{fig:subfigure1}}
%\quad
%\subfigure[]{%
\includegraphics[width=.49\textwidth]{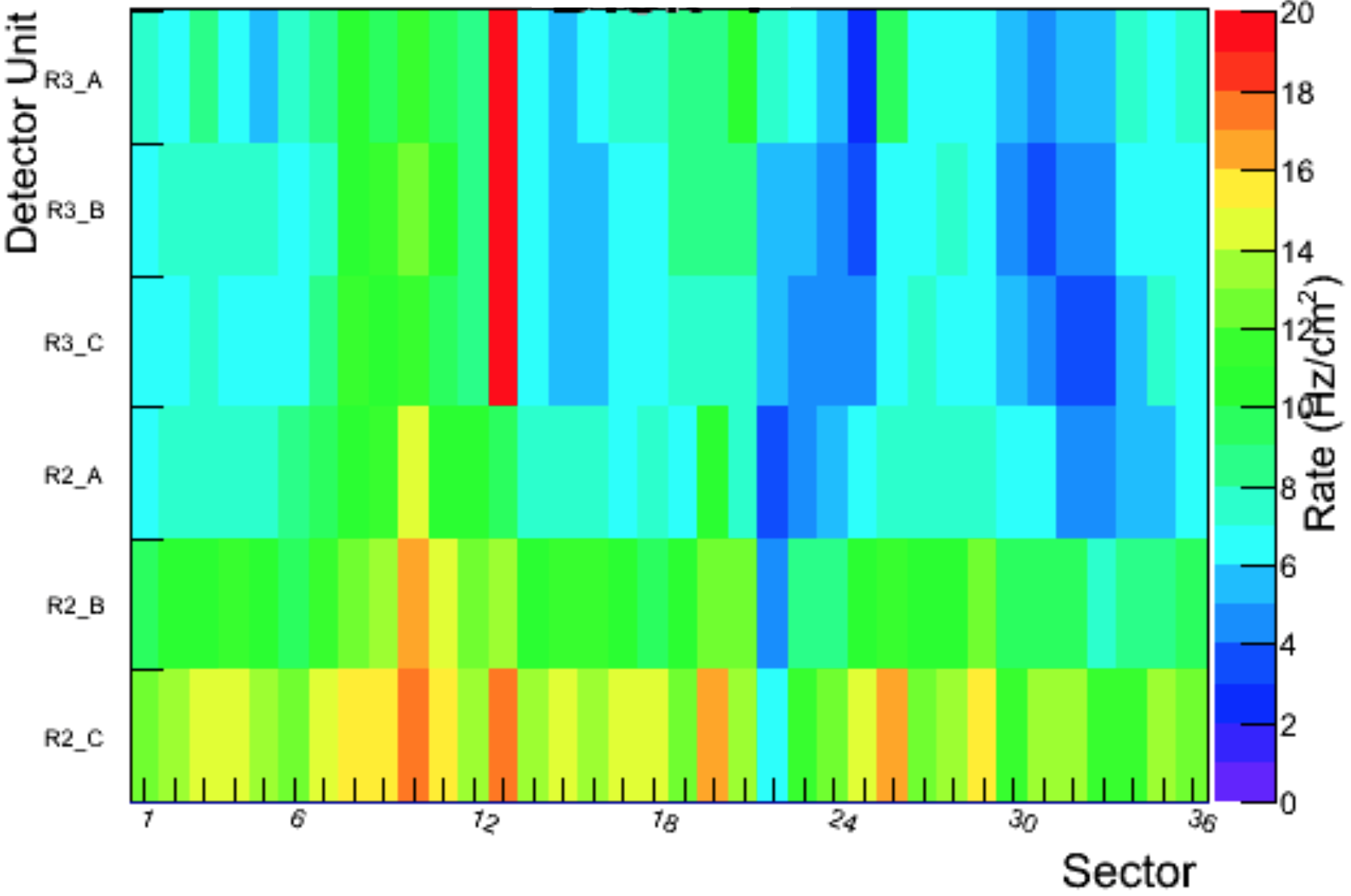}
%\label{fig:subfigure2}}%
\caption{The detector units hit rate (in Hz/cm\textsuperscript{2}) is shown for a run at average instantaneous  luminosity of 4.5*10\textsuperscript{33} cm\textsuperscript{-2}s\textsuperscript{-1}  for the recently installed Disk$\pm4$ in the forward (left) and backward (right) directions during 2015 in at 13 TeV. The X-axis corresponds to the sector, 36 per disk, while the Y-axis corresponds to the partitions of the chamber.}
\label{fig:figure123}
\end{figure}

Figure~\ref{fig:figure123} shows the measured rate for the recently installed Disk$\pm$4, where blue and violet colors correspond to the lower rates, while yellow, orange and red colors correspond to high background level. The average hit rate for these maps is ~10 Hz/cm\textsuperscript{2}. The higher rate for higher eta regions is in agreement with previous measurements and expected from Monte Carlo studies \cite{Chatrchyan:2013sba}.

\begin{figure}[!htb]
\centering
%\subfigure[]{%
\includegraphics[width=.465\textwidth]{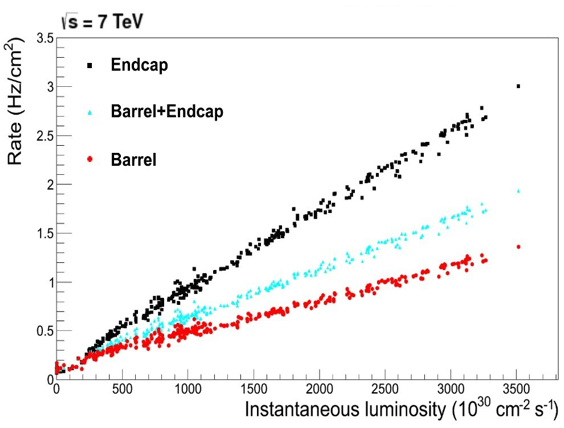}
%\label{fig:subfigure1}}
%\quad
%\subfigure[]{%
\includegraphics[width=.525\textwidth]{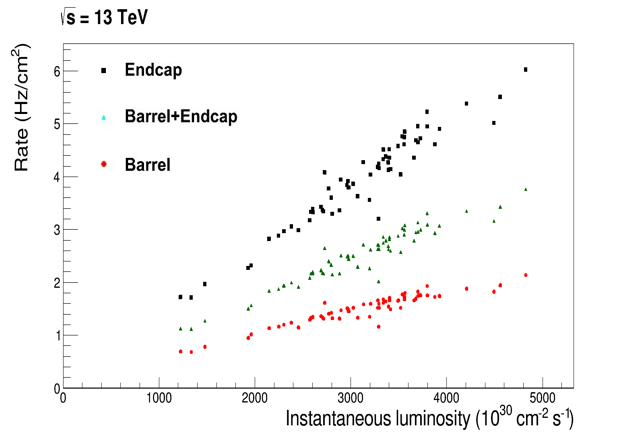}
%\label{fig:subfigure2}}
%
\caption{The plots represent the average hit rate vs. instantaneous luminosity, with 2011 pp collisions at 7 TeV (left) and 2015 pp collisions at 13 TeV (right. The red dots represent the rate measured in barrel and the black represent the rate measured  in endcap. The green markers relate to the overall rate evaluated for the entire RPC system.}
\label{fig:figure1234}
\end{figure}
 
 As shown in figure~\ref{fig:figure1234}, RPC rates increase approximately linearly with the luminosity of LHC. The linear behaviour can be used to extrapolate the rate for future upgrades. 
%///////////////////////////////////////////////////////////////////////////////

\subsection{Mean Cluster Size}

\begin{figure}[!htb]
\centering
%\subfigure[]{%
\includegraphics[width=.485\textwidth]{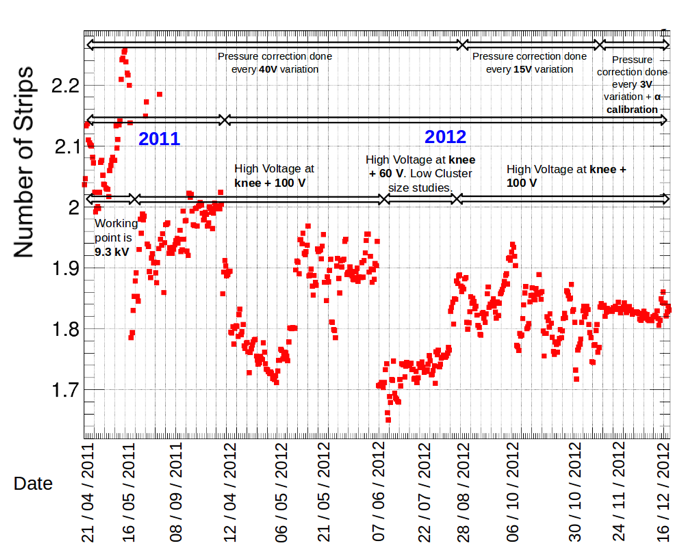}
%\label{fig:subfigure6}}
%\quad
%\subfigure[]{%
\includegraphics[width=.505\textwidth]{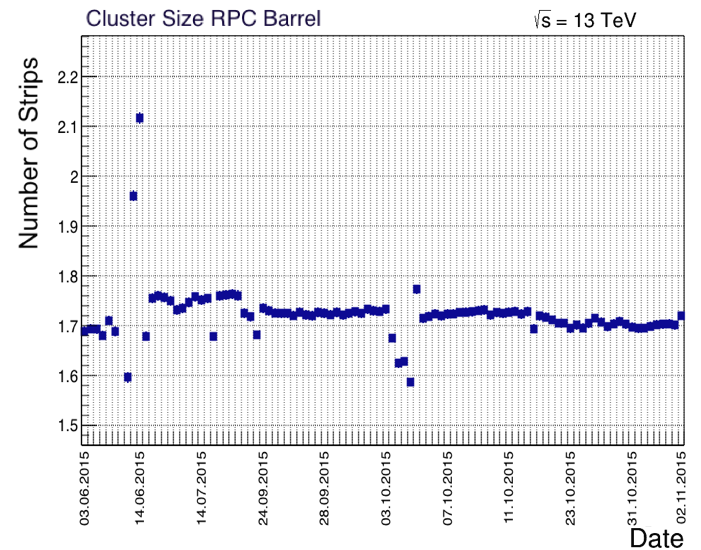}
%\label{fig:subfigure3}}
\caption{The plots represent the history of the mean Cluster Size for the barrel for 2011 and 2012 physics data taking at 8 TeV (left), and for 2015 at 13 TeV (right).}
%\end{figure}
\label{fig:figure1}
\end{figure}

\begin{figure}[!htb]
\centering
%\subfigure[]{%
\includegraphics[width=.49\textwidth]{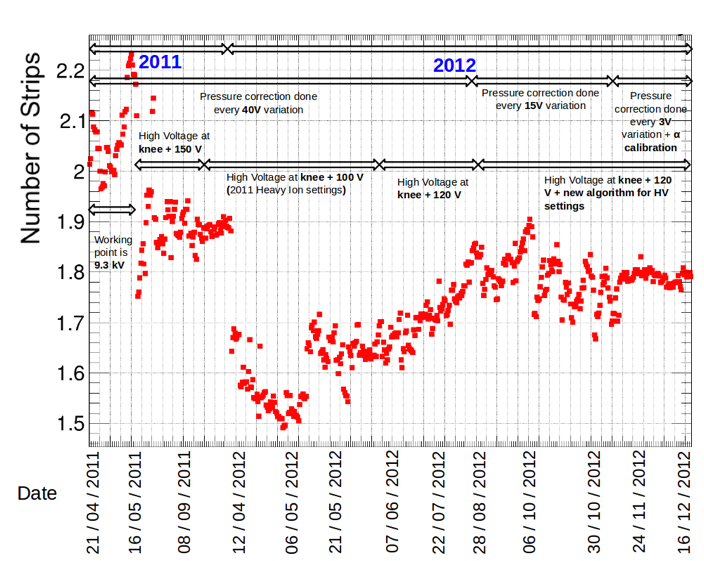}
%\label{fig:subfigure1}}
%\quad
%\subfigure[]{%
\includegraphics[width=.49\textwidth]{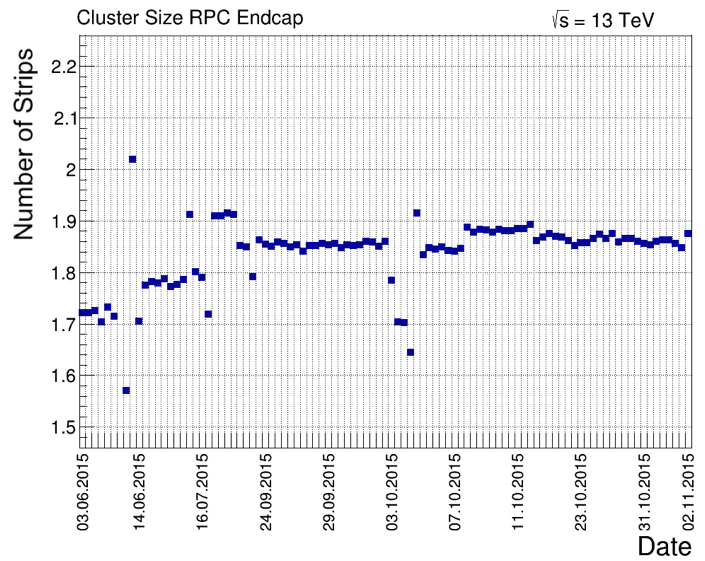}
%\label{fig:subfigure2}}
\caption{The plots represent the history of the mean Cluster Size for the endcap for 2011 and 2012 physics data taking at 8 TeV (left), and for 2015 at 13 TeV (right). }
%\end{figure}
\label{fig:figure99}
\end{figure}

\begin{figure}[!htb]
\centering

\includegraphics[width=.9\textwidth]{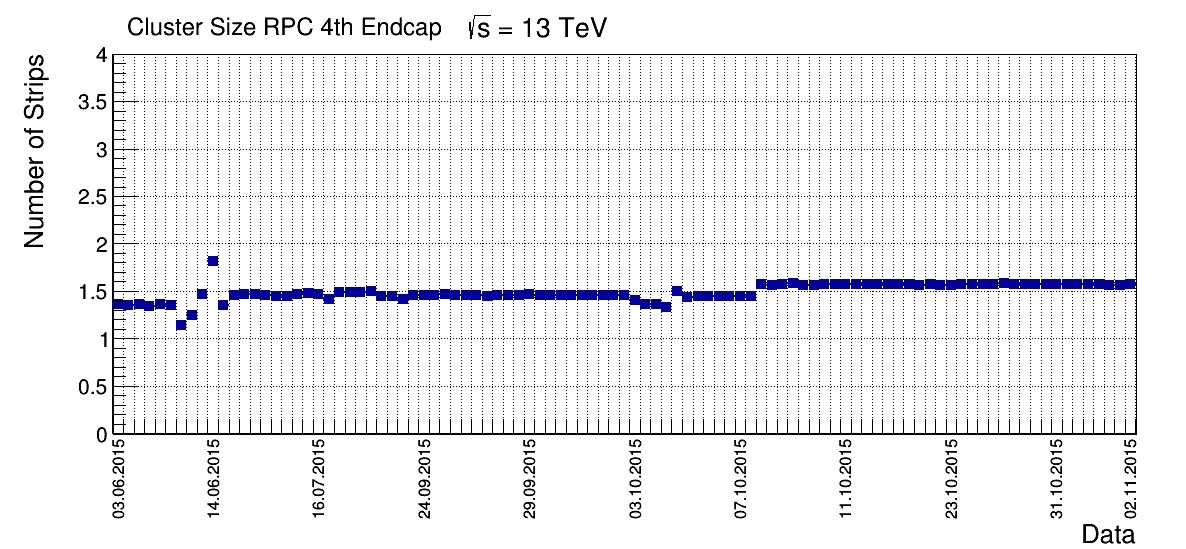}

\caption{The plot represent the history of the mean Cluster Size for Disk$\pm$4 in 2015 at 13 TeV. The fluctuations in the middle of June are due to the performed HV scan. The fluctuations in the beginning of October are due to the performed threshold scan. The increase on the cluster size is due to the change to the optimized WP, applied only to this newly installed stations.}
%\end{figure}
\label{fig:figureM}
\end{figure}

\begin{figure}[htbp]
\centering 
%\subfigure[]{%
\includegraphics[width=.49\textwidth]{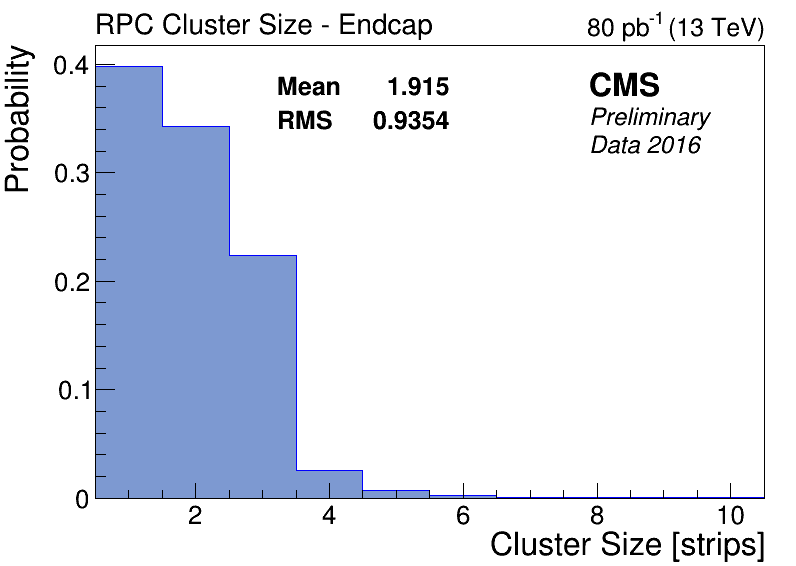} 
%}
%\qquad
%\subfigure[]{%
\includegraphics[width=.49\textwidth]{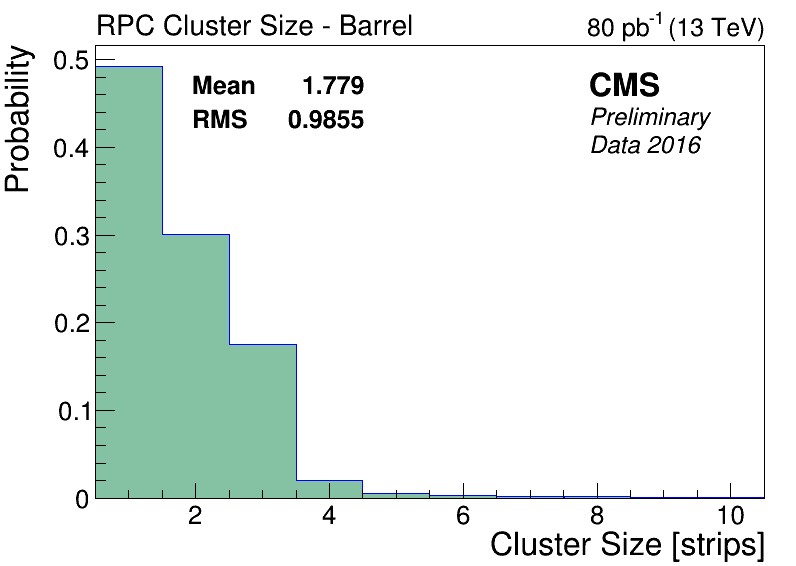}
%}
\caption{\label{fig:cl} Endcap average cluster size (left) and Barrel average cluster size (right).}
\end{figure}

Cluster size (CLS) is defined as the number of adjacent strips fired when an avalanche is produced in the RPC. Keeping the cluster size stable over the time was one of the greatest successes achieved at the end of Run I \cite{Costantini:2012xn}. RPC system has stable average cluster size of about 1.8 strips over the years, which is in agreement with CMS TDR. During RUN2, the chamber cluster size  was monitored run-by-run to guarantie the stability of the system. Average CLS history in 2011 and in start of 2012 is shown in Figure~\ref{fig:figure1} and Figure~\ref{fig:figure99}, is affected by applied pressure corrections and several HV settings. During 2011 and the beginning of 2012 the applied HV to every RPC detector was corrected to compensate for pressure changes in the CMS cavern. The CLS at the end of 2012 was kept lower than 2011 to maintain a stable trigger rate. The fluctuation for 2015 in middle of June and beginning of October, are due to the performed HV and threshold scans respectively. Figure~\ref{fig:figureM} shows the mean CLS history for the newly installed stations in Disk$\pm$4, during 2015 at 13 TeV. Figure~\ref{fig:cl} shows the cluster size distributions for the endcap and barrel measured during 2016. 

%The plot represents the history of the Mean Cluster Size for the newly installed RPC chambers on the 4th Endcap stations. The fluctuations in the middle of June are due to the performed HV scan. The fluctuations in the beginning of October are due to the performed threshold scan. The increase on the cluster size is due to the change of the WP applied to the just to the newly installed Disk-4 (b).

\subsection{Efficiency}
\begin{figure}[htbp]
\centering 
%\subfigure[]{%
\includegraphics[width=.49\textwidth]{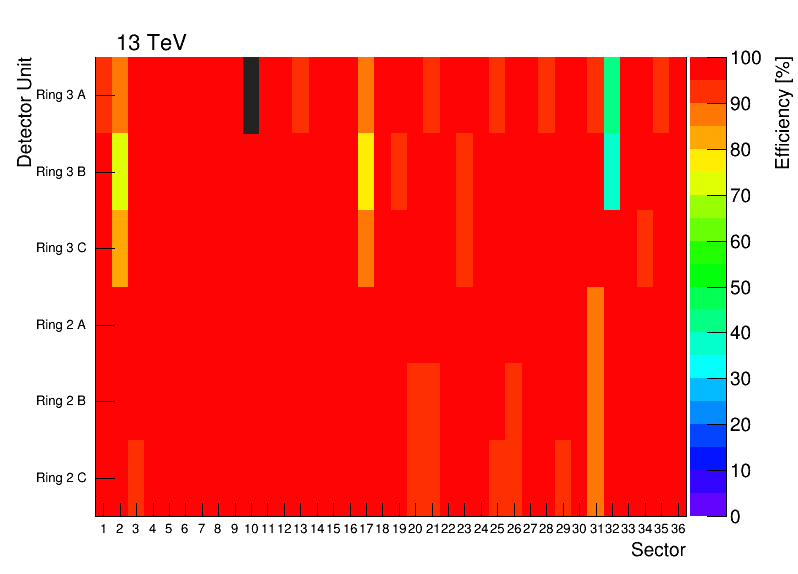}
%}\qquad
%\subfigure[]{%
\includegraphics[width=.49\textwidth]{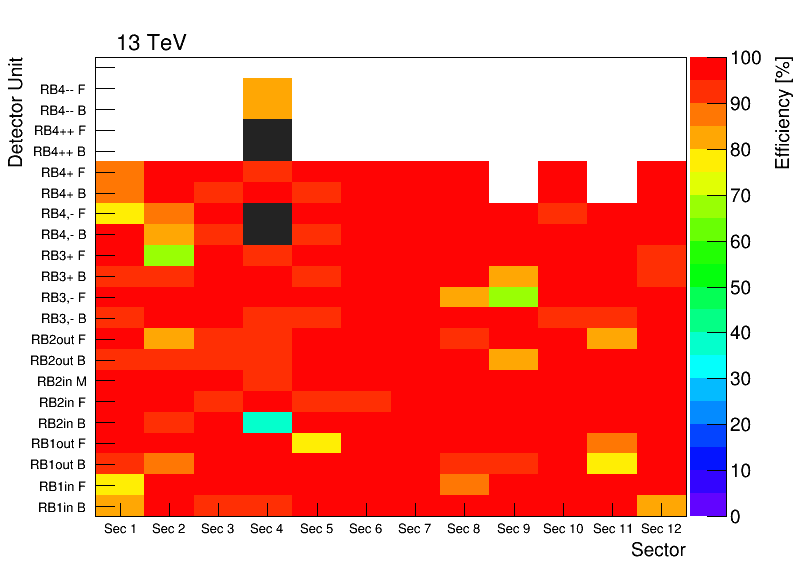}
%}
\caption{\label{fig:eff} Chambers efficiency map for the Disk-1 (left). The 36 sectors are shown on x axis and the 6 $\eta$ partitions on y axis. Chambers efficiency map in Wheel+1 (right). The plot shows the 12 sectors on x axis and the 6 RPC layers (RB1in, RB1out, RB2in, RB2out, RB3, RB4) sections on y axis . For both the black entries correspond to the detector units which are switched off due to known hardware problems. Blue and green colors correspond to the lower efficiency values measured for detector units which have noisy strips that have been masked. }
\end{figure}

Segment extrapolation method ~\cite{Chatrchyan:2013sba} is used to calculate the RPC efficiency. A DT/CSC segment of high quality, associated to a stand-alone muon track, is extrapolated to RPC strip plane. RPC efficiency depends on the atmospheric pressure in the cavern. In order to compensate this dependence, automatic corrections to the HV have been applied during the data taking. As an example, Figure \ref{fig:eff} shows the chambers efficiency map as measured in 2015 for one wheel and for one disk. Most of the chambers have an efficiency of more than 94\%. 

%The inneficient chambers in figure \ref{fig:eff} (a) may be underestimated due to the extrapolation method when the segments are not totally recovered from the CSCs, and it is also responsible for the observed structure as inefficient lines. A new algorithm is 

\begin{figure}[htbp]
\centering 
%\subfigure[]{%
\includegraphics[width=.49\textwidth]{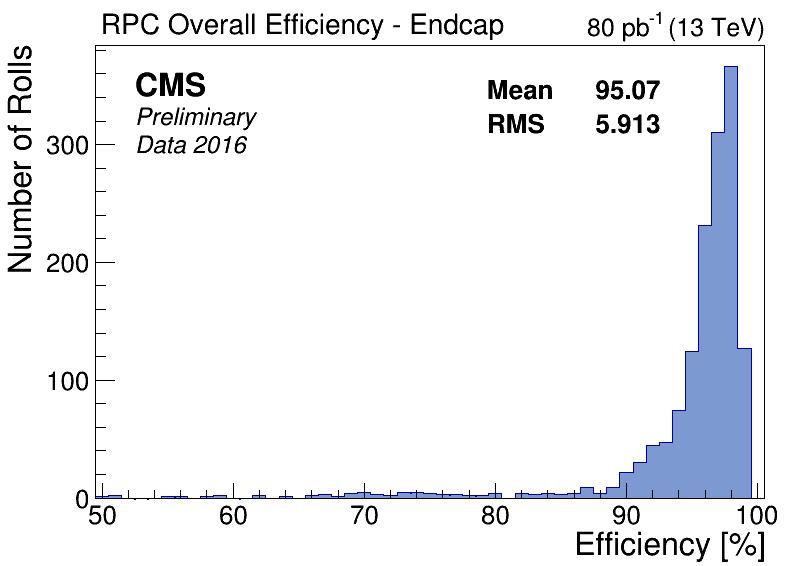}
%}\qquad
%\subfigure[]{%
\includegraphics[width=.49\textwidth]{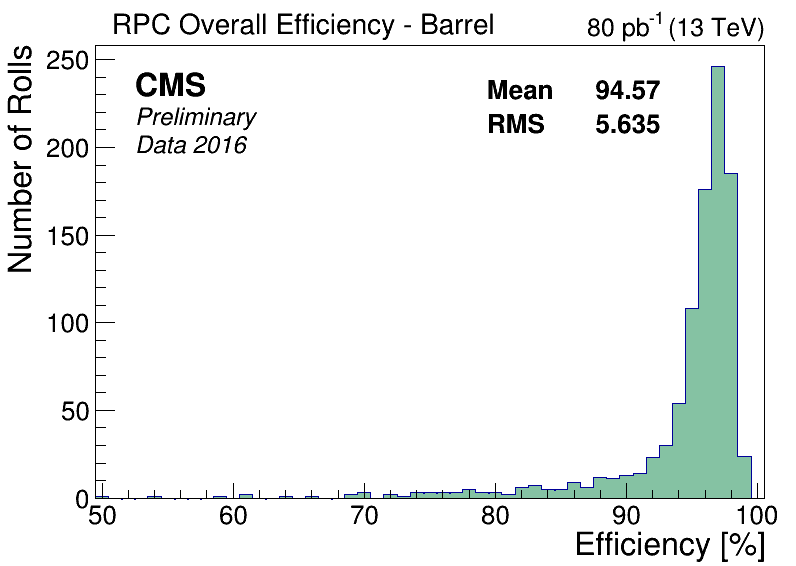}
%}
\caption{\label{fig:eff2016} End-cap chambers efficiency distribution (left) and Barrel chambers efficiency distribution (right). }
\end{figure}

Figure \ref{fig:eff2016} shows the latest measurement of chamber efficiency. The distribution is obtained using 2016 collision data at $\sqrt{s} = 13$  TeV, B = 3.8 T and an integrated luminosity of about 80 pb$^{-1}$. The mean RPC efficiency was calculated to be 94.6 \% - 95.1 \%. The few chambers with low efficiency have known hardware problems. They are propagated in lines because some of the detector units (gaps) share the hardware, like link boards or High Voltage or Low Voltage connections.

\begin{figure}[!htb]
\centering
%\subfigure[]{%
\includegraphics[width=.50\textwidth]{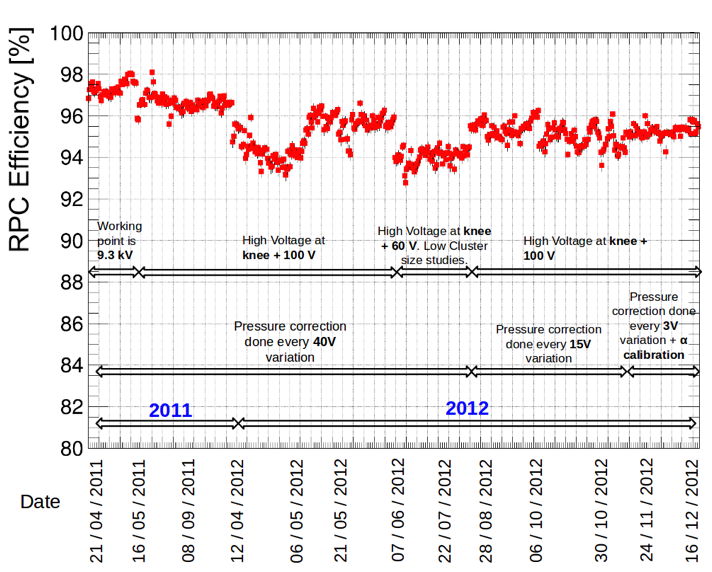}
%\label{fig:subfigure1}}
%\quad
%\subfigure[]{%
\includegraphics[width=.48\textwidth]{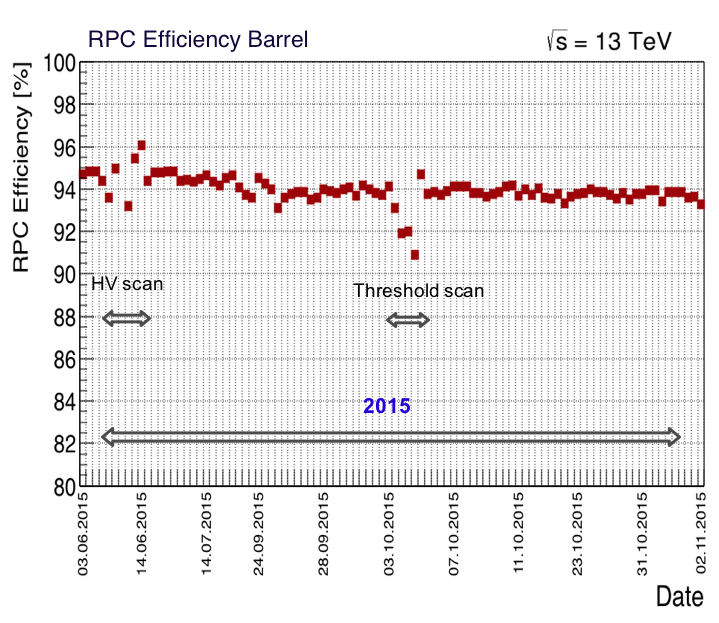}
%\label{fig:subfigure2}}
\caption{The plots represent the history of the overall RPC efficiency for the barrel for 2011 and 2012 physics data taking at 8 TeV (left), and for 2015 at 13 TeV (right.}
\label{fig:figure678}
\end{figure}

\begin{figure}[!htb]
\centering
%\subfigure[]{%
\includegraphics[width=.50\textwidth]{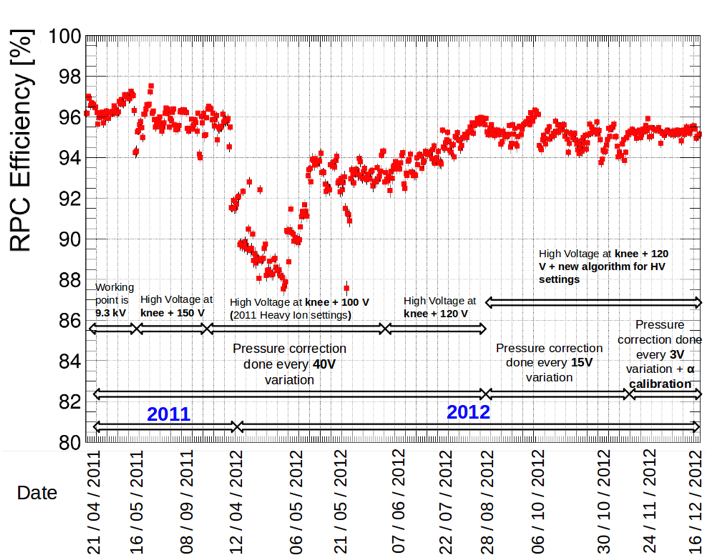}
%\label{fig:subfigure1}}
%\quad
%\subfigure[]{%
\includegraphics[width=.48\textwidth]{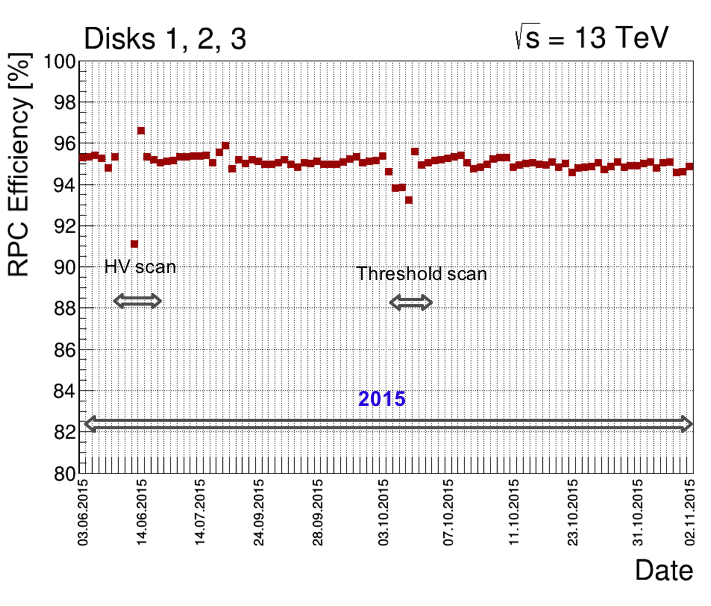}
%\label{fig:subfigure2}}
\caption{The plots represent the history of the overall RPC efficiency for the endcap for 2011 and 2012 physics data taking at 8 TeV in (left), and for 2015 at 13 TeV (right).}
\label{fig:figure66}
\end{figure}

Efficiency is affected  by several HV settings and  applied pressure corrections, during 2011 and beginning of 2012 as shown in figure~\ref{fig:figure678} and figure~\ref{fig:figure66}. The  fluctuation for 2015 in middle of June and beginning of October, are due to the performed HV and threshold scans.

Average RPC efficiency during 2015 at 13 TeV was $\approx$ 94$\%$ after 1 year of LHC running as detectors were operated at lower working points. During  2015 the RPC system was running with a very stable efficiency. The current system works at WP's set at low luminosities, and this may impact the effciency that shows a small degradation at high luminosities.

%\subsubsection{Efficiency of Newly Installed RPC Stations}
%During the first long shutdown (LS1) of the LHC (2013-2014) tfhe CMS muon system was upgraded with 144 RPCs on the endcap (Disk$\pm$4). This addition increased the overall robustness of the CMS muon spectrometer and improved the muon reconstruction efficiency, in the range $1.2 < |\eta| < 1.6$. During the Run II, we put particular attention to these newly installed chambers. 

%The detector performance is monitored via the occupancy distribution of the chambers. The cross-sectional view of both forward and backward stations is shown in Figure~\ref{fig:figure67}. The black points show the position of the reconstructed hits in the middle of the signal electrodes (strips). It is evident from plots that there are no inactive channels in the newly installed chambers.

\begin{figure}[!htb]
\centering
%\subfigure[]{%
%\includegraphics[width=.58\textwidth]{\fig/gg.png}
%\label{fig:subfigure13}}
%\quad
%\subfigure[]{%
\includegraphics[width=.50\textwidth]{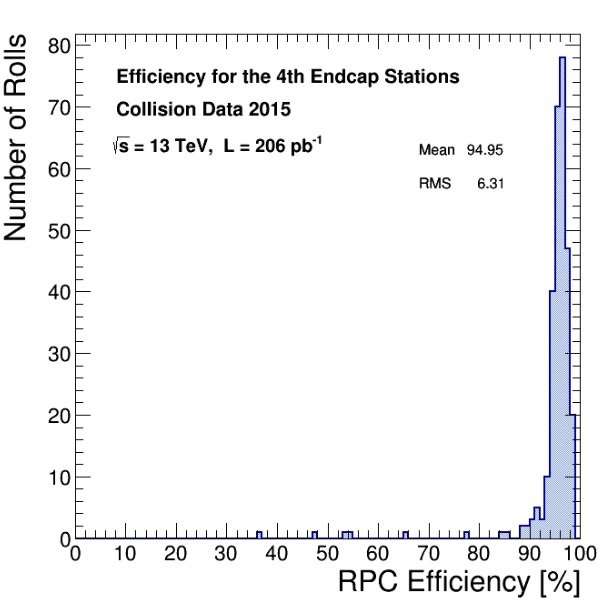}
%\label{fig:subfigure23}}
%\caption{Reconstructed muon hits (XY view occupancy) on the forth positive and negative endcap stations (a) and 
\caption{Overall efficiency distribution of the Disk$\pm$4 stations at 3.8 T}
\label{fig:figure67}
\end{figure}

\begin{figure}[!htb]
\centering
\includegraphics[width=.90\textwidth]{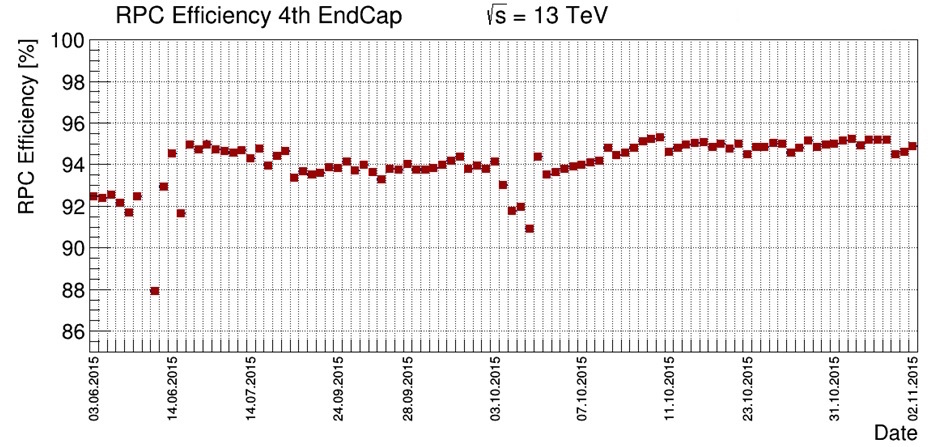}

\caption{History of the overall RPC Efficiency of the newly installed RPCs on the forward and backward endcaps for the 2015 physics data taking.
}
\label{fig:figure68}
\end{figure}

During the Run II, special attention has been put to the newly installed chambers in Disk$\pm$4. Figure \ref{fig:figure67} shows the efficiency distribution for the Disk$\pm$4. The mean value is 94.95\%, which is in good agreement with expectations, and compatible with the rest of the RPC system. The history of the overall RPC efficiency for these chambers over the period of 2015 data taking is shown in Figure~\ref{fig:figure68}.

%Few chambers with low efficiency in the distribution correspond to known hardware problems. 

\section{Intrinsic Noise}

\begin{figure}[htbp]
\centering 
\includegraphics[width=.9\textwidth]{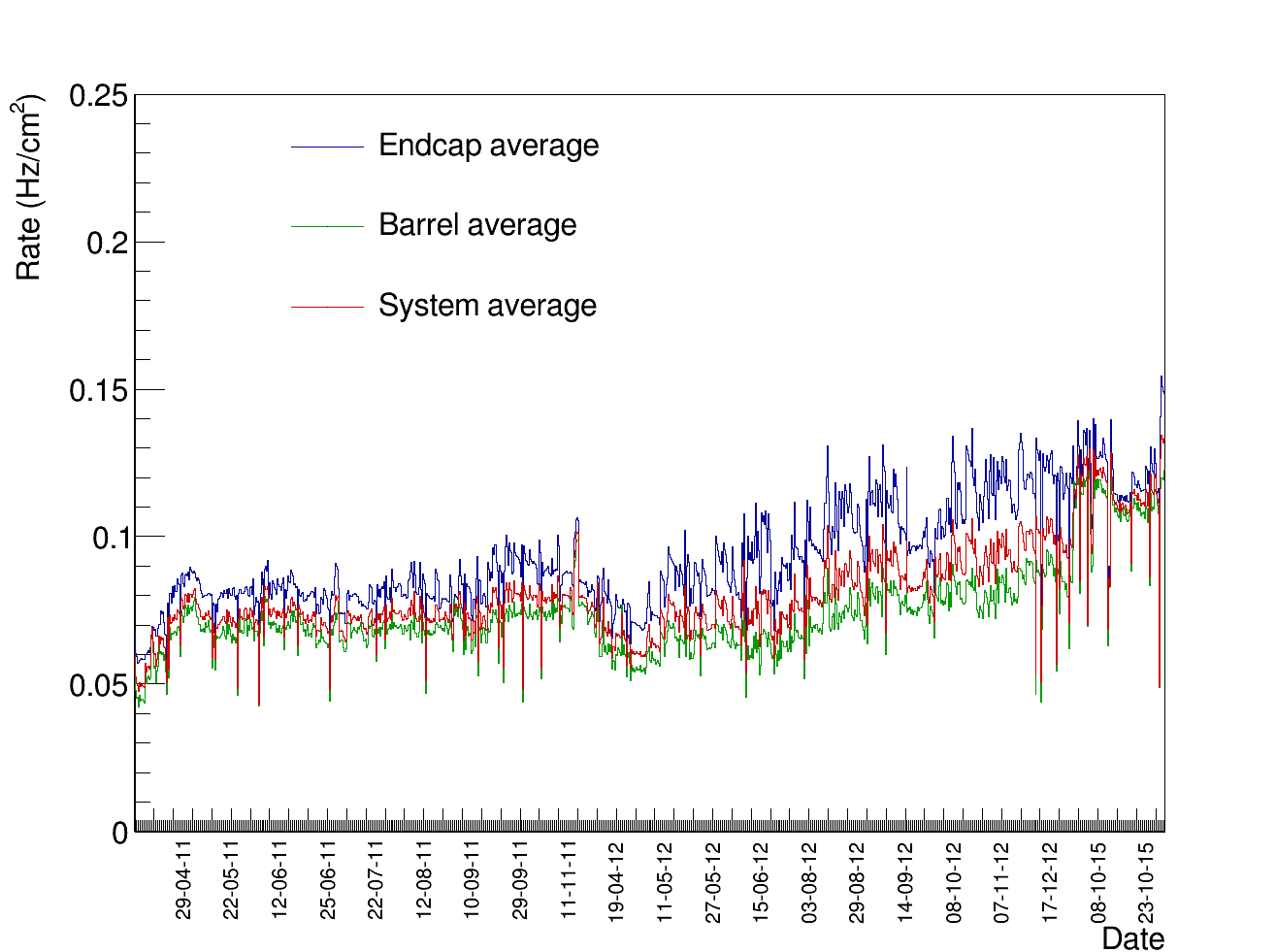}
\caption{\label{fig:noise} RPC cosmic rate distribution.}
\end{figure}

The intrinsic chambers noise and background radiation levels could have an impact on the performance of the system as high rates can affect the trigger performance and the reconstruction of the muon tracks. The RPC rate is measured also during the cosmics data taking in between the collisions runs. Figure \ref{fig:noise} represents the rate level in barrel, endcap and system average from 2011 to 2015. Fluctuations in the rate are mainly due to post$-$collisions radiation, threshold value optimization vs efficiency and operating channels number change. Though the blue and the green curves show similar drift behavior, no significant spike correlations are observed . The overall trend show minor increase in the system rate with time, which is well below the official CMS requirement of rate $<$ 5 $Hz/cm^2$. The end points of the barrel and endcap curves (2015 data taking) get close together since we have lower noise in the Disk$\pm$4 (around 0.05) which lowers the average of the measured endcap rate. 

\begin{figure}[htbp]
\centering 
\includegraphics[width=.9\textwidth]{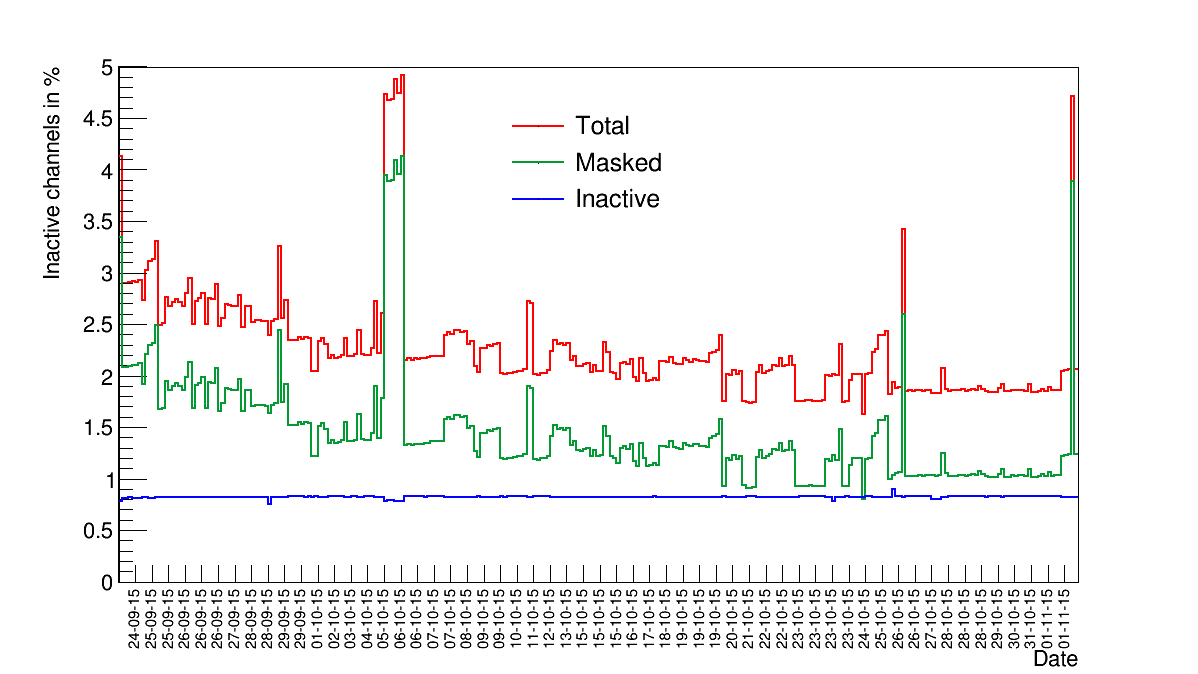}
\caption{\label{fig:innactive} Fraction of channels not operational during 2015.}
\end{figure}

\section{Active Channels}
Before every proton beam fill, the intrinsic noise rate is measured in order to take out of the system (mask) the noisy strips where needed. Figure \ref{fig:innactive} shows the fraction of channels not operational during 2015. The blue line represents the number of inactive (non responsive) channels, while green line represents the number of the masked strips that changes with time as they are adjusted per run depending on the performance of the system. The observed peaks related to the bigger number of masked strips are caused by the temporary hardware problems, which were successfully resolved.In 2015 the percentage of inactive channels was stable between 2 and 2.5\%

\section{Conclusion}

CMS RPC system operated well during RUN2 (2015) delivering good triggers and data for physics. At the end of RUN2, the fraction of active channels was about 98.6\%. Most of inactive channels have been already recovered during LS1. After the LS1 running with increasing instantaneous luminosity and 13 years after the first RPCs have been assembled, the detector performance is within CMS specifications and stable with no degradation observed: average efficiency was found to be about 94\%, average cluster size was  $\sim 1.87$ strips, and the intrinsic noise was $\sim 0.13$ $Hz/cm^2$.

%\section{Sections}
%\subsection{And subsequent}
%\subsubsection{Sub-sections}
%\paragraph{Up to paragraphs.} We find that having more levels usually
%reduces the clarity of the article. Also, we strongly discourage the
%use of non-numbered sections (e.g.~\texttt{\textbackslash
%  subsubsection*}).  Please also see the use of
%``\texttt{\textbackslash texorpdfstring\{\}\{\}}'' to avoid warnings
%from the hyperref package when you have math in the section titles

%\appendix
%\section{Some title}
%Please always give a title also for appendices.

\acknowledgments
We congratulate our colleagues in the CERN accelerator departments for the excellent performance of the LHC and thank the technical and administrative staffs at CERN and at other CMS institutes for their contributions to the success of the CMS effort. In addition, we gratefully acknowledge the computing centers and personnel of the Worldwide LHC Computing Grid for delivering so effectively the computing infrastructure essential to our analyses. Finally, we acknowledge the enduring support for the construction and operation of the LHC and the CMS detector provided by the following funding agencies: BMWFW and FWF (Austria); FNRS and FWO (Belgium); CNPq, CAPES, FAPERJ, and FAPESP (Brazil); MES (Bulgaria); CERN; CAS, MoST, and NSFC (China); COLCIENCIAS (Colombia); MSES and CSF (Croatia); RPF (Cyprus); MoER, ERC IUT and ERDF (Estonia); Academy of Finland, MEC, and HIP (Finland); CEA and CNRS/IN2P3 (France); BMBF, DFG, and HGF (Germany); GSRT (Greece); OTKA and NIH (Hungary); DAE and DST (India); IPM (Iran); SFI (Ireland); INFN (Italy); MSIP and NRF (Republic of Korea); LAS (Lithuania); MOE and UM (Malaysia); BUAP, LNS, CINVESTAV, CONACYT, SEP, and UASLP-FAI (Mexico); MBIE (New Zealand); PAEC (Pakistan); MSHE and NSC (Poland); FCT (Portugal); JINR (Dubna); MON, RosAtom, RAS and RFBR (Russia); MESTD (Serbia); SEIDI and CPAN (Spain); Swiss Funding Agencies (Switzerland); MST (Taipei); ThEPCenter, IPST, STAR and NSTDA (Thailand); TUBITAK and TAEK (Turkey); NASU and SFFR (Ukraine); STFC (United Kingdom); DOE and NSF (USA).

\end{document}